\pdfoutput=1
\documentclass[twocolumn,aps,prd,preprintnumbers,amsmath,amssymb,superscriptaddress,nofootinbib]{revtex4-1}

\usepackage{amssymb,amsmath,bbold}
\usepackage{graphicx,soul}
\usepackage[usenames,dvipsnames]{xcolor}
\usepackage[unicode=true,pdfusetitle,
 bookmarks=true,bookmarksnumbered=false,bookmarksopen=false,citecolor=Turquoise,
 breaklinks=false,pdfborder={0 0 1},backref=false,colorlinks=true,pdfpagemode=FullScreen]
 {hyperref}

\begin{document}

\preprint{IPMU17-0075}
\preprint{CERN-TH-2017-106}

\title{Flavoured $B-L$ Local Symmetry and Anomalous Rare $B$ Decays}

\author{Rodrigo Alonso}
\email[]{rodrigo.alonso@cern.ch}
\affiliation{CERN, Theoretical Physics Department, CH-1211 Geneva 23, Switzerland}
\author{Peter Cox}
\email[]{peter.cox@ipmu.jp}
\affiliation{Kavli IPMU (WPI), UTIAS, University of Tokyo, Kashiwa, Chiba 277-8583, Japan}
\author{Chengcheng Han}
\email[]{chengcheng.han@ipmu.jp }
\affiliation{Kavli IPMU (WPI), UTIAS, University of Tokyo, Kashiwa, Chiba 277-8583, Japan}
\author{Tsutomu T. Yanagida}
\email[]{tsutomu.tyanagida@ipmu.jp}
\affiliation{Kavli IPMU (WPI), UTIAS, University of Tokyo, Kashiwa, Chiba 277-8583, Japan}
\affiliation{Hamamatsu Professor}

\begin{abstract}
We consider a flavoured $B-L$ gauge symmetry under which only the third generation fermions are charged. Such a symmetry can survive at low energies ($\sim\,$TeV) while still allowing for two superheavy right-handed neutrinos, consistent with neutrino masses via see-saw and leptogenesis. We describe a mechanism for generating Yukawa couplings in this model and also discuss the low-energy phenomenology. Interestingly, the new gauge boson could explain the recent hints of lepton universality violation at LHCb, with a gauge coupling that remains perturbative up to the Planck scale. Finally, we discuss more general $U(1)$ symmetries and show that there exist only two classes of vectorial $U(1)$ that are both consistent with leptogenesis and remain phenomenologically viable at low-energies.
\end{abstract}

\maketitle
\section{Introduction}

The Standard Model (SM) with the addition of three right-handed neutrinos provides a very successful model for explaining low-energy observations. 
Small neutrino masses are naturally generated via the seesaw mechanism~\cite{Minkowski:1977sc, *Yanagida:1979as, *Glashow:1979nm, *GellMann:1980vs} and the observed baryon asymmetry is dynamically created through leptogenesis in the early universe~\cite{Fukugita:1986hr}. 
This model also possesses an exact $B-L$ global symmetry in the limit of vanishing Majorana masses for the right-handed neutrinos. 
Thus, following the principle that {\it everything that is allowed is compulsory}, it is natural to promote such a global symmetry to a local one; the Majorana masses would arise in this case from the spontaneous breakdown of the gauged $B-L$ symmetry~\cite{Wilczek:1979hh}. 
The large right-handed neutrino masses required for leptogenesis lead to a very high breaking scale for the $B-L$ symmetry. 
As a consequence, one would not expect to see any effects of the $B-L$ gauge interactions at low energies. 

The above conclusion is however based on the commonly adopted, yet arbitrary, assumption that $B-L$ is generation independent; such an assumption is also unnecessary since gauge anomalies cancel within each generation. 
Furthermore, the generation of neutrino masses and viable leptogenesis both require only two superheavy right-handed neutrinos~\cite{hep-ph/0208157}. 
It is therefore interesting to consider the possibility that a \emph{flavoured} $B-L$ gauge symmetry, under which only the third generation quarks and leptons are charged, could survive at low energies $(\sim\text{TeV})$.
 
In a recent paper~\cite{1704.08158}, we discussed how such a $U(1)_{(B-L)_3}$ symmetry could in fact naturally arise from the breaking of a horizontal $SU(3)_Q\times SU(3)_L\times U(1)_{B-L}$ gauge symmetry at high scales. 
We also pointed out that the resulting low-energy $U(1)_{(B-L)_3}$ gauge boson could explain the recent hints of lepton flavour universality (LFU) violation in rare $B$ decays~\cite{1406.6482, 1705.05802}. The flavour structure of the $U(1)_{(B-L)_3}$ allows it to naturally evade otherwise fatal bounds from FCNCs involving the first two generations. At the same time, constraints from LHC searches allow for masses as light as a TeV, as opposed to previous models.

The purpose of this letter is to describe this model in detail. 
We discuss how the Yukawa couplings of the three known families of quarks and leptons can be generated via the addition of a single family of vector-like fermions, with masses of order the $U(1)_{(B-L)_3}$ breaking scale. 
We also show explicitly that the model can indeed explain the observed $B$-decay anomalies without conflicting with existing experimental results, and while remaining perturbative and self-consistent up to the Planck scale.

\section{Model}
We assume that only the third generation of fermions is charged under the local $U(1)_{(B-L)_3}$, such that the charges read, in flavour space,
\begin{align} \label{Torig}
  T^{q}&=\frac13
  \left(\begin{array}{ccc}
  0&0&0\\
  0&0&0\\
  0&0&1
  \end{array}\right),&
  T^{l}&=
  \left(\begin{array}{ccc}
  0&0&0\\
  0&0&0\\
  0&0&-1
  \end{array}\right) ,
\end{align}
while being vectorial, i.e. the same for LH and RH fields. The SM Higgs $H$ is assumed to be neutral under $U(1)_{(B-L)_3}$. This symmetry does not allow for couplings between the third and the first two generations, hence, the Yukawa couplings are
\begin{align} 
  \mathcal{L}_{Y_f}=& -\bar{q}_{L} \hat Y_u  \tilde{H} u_R-  \bar{q}_{L} \hat Y_d H d_R -  \bar{l}_{L}  \hat Y_eH e_R- \bar{l}_{L} \hat Y_\nu \tilde{H}\nu_R\nonumber \\
  & -\bar{\mathrm{q}}_{L}^{\,3} Y_t  \tilde{H} t_R-  \bar{\mathrm{q}}_{L}^{\,3}  Y_b H b_R -  \bar{\mathrm l}_{L}^{\,3}   Y_\tau H \tau_R- \bar{\mathrm{l}}_L^{\,3} Y_{\nu_3} \tilde{H}\mathrm{\nu}_R^{\,3}\nonumber \\
  &-\frac{1}{2} \bar{\nu}^c_R \hat M_{\nu_R} \nu_R+h.c. \,, \label{L-Yf}
\end{align}
where $\tilde{H}=i\sigma_2 H^*$ and we have separated the first and second generation fields $q_L\,u_R\,,d_R\, l_L\,e_R\, \nu_R$, with an implicit index that runs from 1 to 2 (e.g. $d_R=(d_{R_1}\,d_{R_2})$), from the third generation fields $\mathrm{q}_L^{\,3}\,, b_R\,, t_R\,, \mathrm{l}_L^3 \,, \tau_R\,,\nu_R^3$. 
Yukawa couplings with an upper wedge $\hat Y$ are $2\times 2$ matrices, whereas those without and with a 3rd family subindex are constants. 
This is best visualized in matrix form:
\begin{align} \label{YukStr0}
  Y_{d}=
  \left(\begin{array}{cc}
  \hat Y_d &0\\
  0 &Y_b
  \end{array}\right),\quad
  M_{\nu_R}=
  \left(\begin{array}{cc}
  \hat M_{\nu_R}&0\\
  0&0
  \end{array}\right) ,
\end{align}
where $Y_d$, $M_{\nu_R}$ are the usual $3\times 3$ Yukawa couplings and Majorana masses and similar Yukawa expressions hold for up-type quarks and leptons.

The above Yukawa structure does not lead to mixing between the third generation and the first two, so a mechanism should be put in place to `fill in the zeros' in Eq.~(\ref{YukStr0}). 
This is done here by introducing scalar fields with $U(1)_{(B-L)_3}$ charge $\phi_q(+\frac{1}{3})$, $\phi_l(+1)$, $\chi(+2)$, which also do the job of breaking the $U(1)$ symmetry, and a vector-like fermion for each spin $1/2$ representation of the SM gauge group, $Q_{L,R}, U_{L,R}, D_{L,R},L_{L,R},E_{L,R}\,N_{R,L}$, which are neutral under $U(1)_{(B-L)_3}$.\footnote{Alternatively, one can introduce four new Higgs doublets which carry $U(1)_{(B-L)_3}$ charges $\pm\frac{1}{3}$ and $\pm 1$. However, these can lead to potentially dangerous FCNCs.} 
However, it should be noted that not all vector-like fermions are required to generate the observed masses and mixings.

\begin{figure}[htbp]
  \centering
  \includegraphics[width=0.5\textwidth]{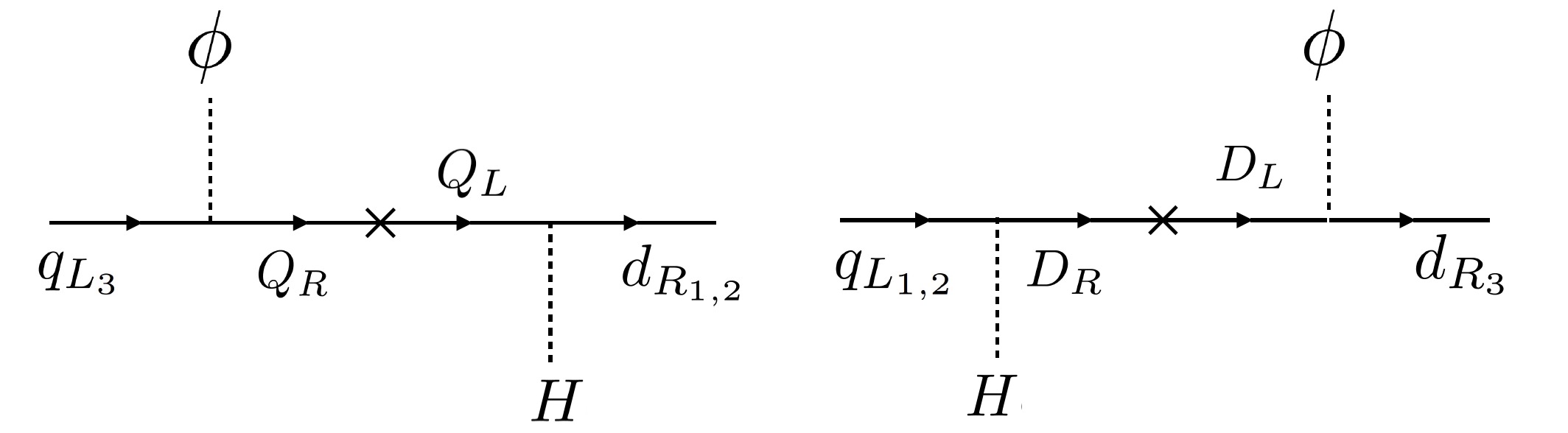} 
  \caption{Illustration of fermion mass generation.}
  \label{feynman}
\end{figure}

The most general renormalisable Lagrangian then reads, in the quark sector and in addition to Eq.~(\ref{L-Yf}):
\begin{align}
  \mathcal{L}_{Y_Q}= &- \bar{q}_{L} Y_D  H D_R -Y_D^\prime\phi_q^* \bar{D}_L b_{R}-M_D \bar{D} D_R \nonumber \\
  &- \bar{q}_{L} Y_U H U_R-Y_U^\prime\phi_q^* \bar{U}_L t_{R}-M_U \bar{U} U_R \nonumber \\
  &-Y^\prime_Q\phi_q\, \bar{\mathrm q}_{L}^{\,3} Q_R- \bar{Q}_{L} H Y_Q^T d_{R}- \bar{Q}_{L} \tilde{H} \tilde{Y}_Q^Tu_{R}  \nonumber \\ 
  &-M_Q \bar{Q}_L Q_R+h.c. \,, \label{L-YQ}
\end{align}
where $Y_{Q}, \tilde{Y}_{Q}, Y_{U,D}$ are 2-vectors, and the rest are complex constants. 
The lepton sector has the same structure but in addition we have:\footnote{If we do not introduce $\chi$, a higher dimensional operator could generate the Majorana mass. Since the mass is suppressed, the third right-handed neutrino could be identified with dark matter.}$^,$\footnote{For simplicity, we assume vanishing Majorana masses for $N_{L,R}$.}
\begin{align}
  - \bar \nu_R^c \lambda_\phi \phi_l \nu_R^3- \frac12\lambda_\chi \chi \bar\nu_{R}^{3c}\nu_R^3+h.c \,.
\end{align}
where $\lambda_\phi$ is a 2-vector and $\lambda_\chi$ a complex constant.
The vector-like fermions are assumed to be much heavier than the SM fields ($M_{Q,D\,...}\gg\left\langle H\right\rangle$) and so can be integrated out. Taking the down quark mass matrix as an example, we obtain
\begin{align}\label{YdMass}
  \left(\begin{array}{cc}\bar q_{L_{1,2}}& \bar {\mathrm{q}}_{L}^3\\\end{array}\right)
  H \left(\begin{array}{cc}
  \hat Y_d   &    -\frac{Y^\prime_D \phi_q^*}{M_D} Y_D    \\
  - \frac{Y^\prime_Q \phi_q}{M_Q} Y_Q^T       & Y_b\\
  \end{array} 
  \right)
  \left(\begin{array}{c} d_{R_{1,2}}\\ b_{R}\\\end{array}\right) ,
\end{align}
and similarly for up-type quarks, and charged and neutral leptons. 
The mechanics is shown in Fig.~\ref{feynman} diagramatically.

Neutrinos still get their mass through the usual seesaw mechanism:
\begin{align} \label{nuLHMass}
  \frac{1}{2} \left(\begin{array}{cc} \bar l_{L_{1,2}}^c & \bar{\mathrm l}_L^{3c}\end{array} \right)\tilde H^T Y_\nu^* \frac{1}{M_{\nu_R}} Y_\nu^\dagger \tilde H\left(\begin{array}{c}l_{L_{1,2}}\\\mathrm{l}_L^{\,3}
  \end{array}\right)+h.c. \,,
\end{align}
where
\begin{align}
  \frac{1}{M_{\nu_R}}&\simeq
  \left(\begin{array}{cc}
  \frac{1}{\hat M_{\nu_R}}&-\frac{\phi_l}{\lambda_\chi\chi \hat M_{\nu_R}}\lambda_\phi\\
  -\lambda_\phi^T\frac{ \phi_l}{\lambda_\chi\chi \hat M_{\nu_R}}&\frac{1}{\lambda_\chi\chi}\\
  \end{array}\right),\\
  Y_\nu&=\left(\begin{array}{cc}
  \hat Y_\nu   &    -\frac{Y^\prime_N \phi_l^*}{M_N} Y_N    \\
  - \frac{Y^\prime_L \phi_l}{M_L} Y_L^T  & Y_{\nu_3}\\
  \end{array} 
  \right) .
\end{align}
Given the hierarchy $\hat M_{\nu_R}\gg \chi$, the $1/\chi$ entry in $M_{\nu_R}^{-1}$ is much larger than the rest, which implies that to get the correct LH neutrino mass scale $Y_{\nu_3}\,, Y_N  Y_N^\prime \phi_l/M_N \lesssim 10^{-5}$.

The final step is to diagonalize the mass matrices obtained via the above mechanism; we have a unitary rotation for each field $f=U_f f^\prime$ such that:
\begin{align}\label{Masses}
  U_{d_L}^\dagger Y_{d} U_{d_R}&={\rm{diag}}(m_d,m_s,m_b)\sqrt{2}/v \,, \\ \nonumber
  U_{u_L}^\dagger Y_{u} U_{u_R}&={\rm{diag}}(m_u,m_c,m_t)\sqrt{2}/v \,, \\  \nonumber
  U_{e_L}^\dagger Y_{e} U_{e_R}&={\rm{diag}}(m_e,m_\mu,m_\tau)\sqrt{2}/v \,, \\  \nonumber
  U_{\nu_L}^T Y_\nu^* M_{\nu_R}^{-1}Y_\nu^\dagger U_{\nu_L}&={\rm{diag}}(m_{\nu_1},m_{\nu_2},m_{\nu_3})\,2/v^2 \,,
\end{align}
and we recall that $U_{u_L}^\dagger U_{d_L}=V_{CKM}$ and $U_{e_L}^\dagger U_{\nu_L}=U_{PMNS}$. 
These relations together with those in Eq.~(\ref{Masses}) comprise the known values of the flavour structure that the Yukawas generated as in Eqs.~(\ref{YdMass}) have to reproduce.
It is clear, since general $3\times 3$ Yukawa couplings have been generated, that these conditions can be satisfied with unconstrained parameters remaining in the model.

Here, for definiteness, we adopt a well-motivated simplifying ansatz regarding the free parameters in the unitary matrices; details about how to obtain this structure can be found in appendix~\ref{appendix}. 
Firstly, we take $M_Q\gg M_{U,D}$ and $M_L\gg M_E$, which is a limit in which rotations of the third generation RH charged fermions are highly suppressed. 
As for the mixing induced by $U,D$ and $E$ in the LH third family fields, we allow for the generation of two additional angles, $\theta_{l,q}$, beyond those present in $V_{CKM} $ and $U_{PMNS}$. 
For phenomenological reasons, we assume both of these angles correspond to a $2-3$ family rotation.
These assumptions made explicit read:
\begin{align}
  U_{e_L}&=R^{23}(\theta_l),&  U_{\nu_L}&=R^{23}(\theta_l)U_{PMNS}, \nonumber \\
  U_{d_L}&=R^{23}(\theta_q),  &U_{u_L}&=R^{23}(\theta_q)V^\dagger_{CKM} ,  \label{Rots}
\end{align}
where $R^{ij}(\alpha)$ is a rotation in the $ij$ sector by an angle $\alpha$. The connection of these rotation matrices to the model parameters in Eq.~(\ref{L-YQ}) is deferred to appendix~\ref{appendix}.

Finally, the new gauge boson, $Z_{BL3}$, interacts with SM fields according to
\begin{equation}
  \mathcal{L}_{Z_{BL3}} = \frac12Z_{BL3}^\mu \left(\partial^2+M^2 \right)Z_{BL3,\mu}-g Z^\mu_{BL3} J_\mu \,,
\end{equation}
where
\begin{equation}
  J_\mu=\sum_f\bar{f} U^\dagger_f T^f U_f \gamma_\mu f \,,
\end{equation}
with $T^f$ as given in Eq.~(\ref{Torig}) and unitary rotations as in Eq.~(\ref{Rots}).

\section{Low energy phenomenology} \label{sec:pheno}
The most significant low-energy consequences of this model are in flavour observables, particularly FCNC processes mediated by the $Z_{BL3}$.
While the $Z_{BL3}$ may also be directly produced at the LHC, the suppressed couplings to first and second generation quarks mean that the bounds are significantly weaker than in generic $Z'$ models. 
We discuss these constraints in detail below, focusing on $Z_{BL3}$ masses $\gtrsim\,$TeV. 
In this mass range, effects in other low-energy observables such as neutrino scattering and $(g-2)_\mu$ are safely below existing bounds.
Lastly, there will be $Z-Z_{BL3}$ kinetic mixing via the Lagrangian term $\epsilon \tilde{F}F_{BL3}$, where $\epsilon$ is a free parameter. 
For $M\sim\,$TeV, the constraints are relatively weak, $\epsilon\lesssim0.4$~\cite{1006.0973}.

\subsection{Semi-leptonic $B$ Decays}

There has recently been significant interest in hints of LFU violation in semi-leptonic $B$ decays, as observed by LHCb~\cite{1406.6482, 1705.05802}. 
Measurements of the ratios
\begin{equation} \label{R_K}
  \mathcal{R}_K^{(*)}=\frac{\Gamma\left(B\rightarrow K^{(*)}\mu^+\mu^-\right)}{\Gamma\left(B\rightarrow K^{(*)}e^+e^-\right)} \,,
\end{equation}
show a consistent departure from the SM prediction, which is under excellent theoretical control~\cite{hep-ph/0310219}. 
In fact, global fits to the data suggest significant tension with the SM at around the $4\sigma$ level~\cite{1704.05435,1704.05438,1704.05340,1704.05444,1704.05447,Geng:2017svp}. 

The relevant effective Hamiltonian {involving charged leptons  that will receive contributions from  $Z_{BL3}$ is defined as
\begin{equation}
  \mathcal{H}_{\text{eff}} = -\frac{4G_F}{\sqrt{2}}V_{tb}V_{ts}^*\left(C_9^l\mathcal{O}_9^l+C_{10}^l\mathcal{O}_{10}^l  \right) \,,
\end{equation}
with
\begin{align}
  \mathcal{O}_9^l&=\frac{\alpha}{4\pi}\left(\bar s\gamma_\mu \,b_L\right)\left(\bar l\gamma_\mu l \right)\,, \\ 
  \mathcal{O}_{10}^l&=\frac{\alpha}{4\pi}\left(\bar s\gamma_\mu \,b_L\right)\left(\bar l\gamma_\mu\gamma^5 l \right)\,.
\end{align}
It is well-known that a significantly improved fit to the data can be obtained by an additional contribution to the Wilson coefficients $C_9$ and $C_{10}$. 
In our model\footnote{
For other $Z^\prime$ explanations of the anomaly see~\cite{1310.1082, *1311.6729, *1403.1269, *1501.00993, *1503.03477, *1503.03865, *1505.03079, *1506.01705, *1507.06660, *1509.01249, *1510.07658, *1511.07447, *1601.07328, *1604.03088, *1608.01349, *1608.02362, *Crivellin:2016ejn, *1701.05825, *1703.06019, *1704.06005}.}, integrating out the $Z_{BL3}$ yields the effective Lagrangian
\begin{equation}
  \mathcal{L}=-\frac{g^2s_{\theta_q}c_{\theta_q}s_{\theta_l}^2}{3M^2} (\bar{s} \gamma^\rho b_L)(\bar{\mu} \gamma_\rho \mu_L) \,,
\end{equation}
which results in a contribution
\begin{equation} \label{eq:C9}
  \delta C_9^\mu=-\delta C_{10}^\mu= -\frac{\pi}{\alpha\sqrt{2} G_F V_{tb} V^*_{ts}}\frac{g^2s_{\theta_q}c_{\theta_q}s_{\theta_l}^2}{3M^2}\,.
\end{equation}
The best fit-region to the data (assuming $\delta C_9^\mu=-\delta C_{10}^\mu$) is given by $\delta C_9^\mu\in[-0.81\,-0.48]$ $([-1.00,\,-0.34])$ at $1(2)\sigma$~\cite{1411.4413}. 
In order to explain the LFU anomalies we therefore require $\theta_q<0$ (for small $\theta_q$). Unlike in other models (e.g.~\cite{1704.08158}), this precludes the simple possibility that the rotation in the down sector is given by the CKM, i.e. $U_{d_L}\neq V_{CKM}$. The best-fit region is shown in Fig.~\ref{fig:M-theta_q}.

Note that the above Wilson coefficients also contribute to the fully leptonic decay $B_s\rightarrow\mu\mu$, however the best-fit region is consistent with the existing measurements. 
$SU(2)_L$ gauge invariance also ensures that there is a similar contribution to the decays $B\rightarrow K^{(*)}\nu\bar{\nu}$, although this results in only sub-dominant constraints on the parameter space.

\subsection{Meson Mixing}

The strongest constraints on this model come from contributions to the mass difference in $D^{0}-\bar{D^{0}}$ and, in particular, $B_s-\bar{B}_s$ mixing. 
The relevant effective Lagrangian is
\begin{equation}
  \mathcal{L}=-\frac{g^2s_{\theta_q}^2c_{\theta_q}^2}{18M^2} (\bar{s} \gamma^\mu b_L)^2 -\frac{g^2c_D^2}{18M^2} (\bar{u} \gamma^\mu c_L)^2 \,,
\end{equation}
where:
\begin{equation}
  c_D\equiv\left(V_{ub}\,c_{\theta_q}-V_{us}\,s_{\theta_q}\right)\left(V^*_{cb}\,c_{\theta_q}-V^*_{cs}\,s_{\theta_q}\right) \,.
\end{equation}
This leads to 
\begin{equation}
  C_{B_s}\equiv\frac{\Delta m_{B_s}}{\Delta m_{B_s}^\text{SM}}=1+\frac{4\pi^2c(M)}{G_F^2m_W^2V_{tb}V^*_{ts}\,\hat{\eta}_BS(\frac{m_t^2}{m_W^2})}\frac{g^2s_{\theta_q}^2c_{\theta_q}^2}{18M^2} \,,
\end{equation}
and
\begin{equation} \label{eq:D_mixing}
  \Delta m_D^{NP}=\frac{2}{3}f_D^2 B_D m_D\, c(M) \frac{g^2c_D^2}{18M^2}\,.
\end{equation}
The factor $c(M)\approx0.8$ includes the NLO running~\cite{hep-ph/9711402, hep-ph/0005183} down to the meson mass scale. 
For the $B_s$ system, the SM prediction is given in terms of the Inami-Lim function $S(m_t^2/m_W^2)\approx2.30$~\cite{Inami:1980fz}, and $\hat{\eta}_B\simeq0.84$ accounts for NLO QCD corrections~\cite{Buras:1990fn,1008.1593}.
Measurements of the mass difference result in the stringent constraint $0.899<C_{B_s}<1.252$ at 95\% CL~\cite{0707.0636}.

In the case of $D^{0}-\bar{D^{0}}$ mixing, the SM prediction suffers from significant uncertainties~\cite{0705.3650} and we simply require that the contribution in Eq.~\eqref{eq:D_mixing} not exceed the measured value, $0.04<\Delta m_D<0.62$ at 95\% CL~\cite{1612.07233}. 
We use the lattice values $f_D=207.4\,$MeV~\cite{1411.7908} and $B_D=0.757$~\cite{1505.06639}.

The strong bounds from, in particular, $B_s-\bar{B}_s$ mixing can be clearly seen in Fig.~\ref{fig:M-theta_q}. 
Nevertheless, for a sufficiently small mixing angle, $|\theta_q|\lesssim0.15$, the LFU anomalies can be explained while remaining consistent with the current bounds. 
Note that, for a given value of $\theta_l$, this upper limit on the mixing angle is determined solely by the ratio of the $U(1)_{(B-L)_3}$ charges in the quark and lepton sectors. 
Finally, decreasing the mixing angle in the lepton sector reduces the contribution to $\delta C_9^\mu$, meaning a smaller $|\theta_q|$ is required in order to simultaneously satisfy the bounds from meson mixing.

\subsection{Lepton Flavour Violation}

Depending on the mixing angle in the lepton sector, the $Z_{BL3}$ may also mediate lepton flavour violating processes. 
In particular the decay $\tau\to3\mu$, which is tightly constrained by experiment: $\text{BR}(\tau\rightarrow 3\mu)<2.1\times10^{-8}$ at 90\% CL~\cite{1001.3221}. 
The effective Lagrangian
\begin{align}
  \mathcal{L}_{\text{LFV}} = \frac{g^2}{M^2}s^3_{\theta_l} c_{\theta_l}\, \bar \tau\gamma^\rho \mu_L \,\bar\mu\gamma_\rho \mu_L \,,
\end{align}
results in a branching ratio
\begin{equation}
  \text{BR}(\tau\rightarrow 3\mu)=\frac{m_\tau^5}{1536\pi^3\Gamma_\tau}\frac{2g^4}{M^4}s^6_{\theta_l} c^2_{\theta_l} \,.
\end{equation}
The experimental bounds can be trivially satisfied for $\theta_l\approx\pi/2$, but already disfavour maximal mixing ($\theta_l\approx\pi/4$) if one wishes to simultaneously explain the anomalies. 
The $Z_{BL3}$ can also mediate the LFV decay $B\to K^{(*)}\tau\mu$, although the branching ratio lies well below current experimental bounds~\cite{1504.07928}.

\subsection{Collider Searches}
The fact that the $Z_{BL3}$ only couples to third generation quarks (in the flavour basis), ensures that its production cross section at the LHC is significantly suppressed compared to a generic $Z'$ with flavour universal couplings. 
After rotating to the mass basis there will be couplings to, in particular, the second generation quarks. 
However, constraints from $B_s-\bar{B}_s$ mixing already require that the mixing angle $\theta_q$ is relatively small, such that $b\bar{b}\to Z_{BL3}$ remains the dominant production channel. 
The LHC bounds are then effectively independent of $\theta_q$ in the relevant region of parameter space. 
Furthermore, regions of parameter space which can explain the LFU anomalies will have a sizeable branching ratio to muons, making this the most promising search channel. Alternatively, in the event of negligible mixing in the charged lepton sector, $t\bar{t}$, $b\bar{b}$ and $\tau\bar{\tau}$ resonance searches can be used, but yield significantly weaker bounds.

In Fig.~\ref{fig:M-g} we show the bounds from the latest ATLAS di-muon search with 36$\,\text{fb}^{-1}$ at $\sqrt{s}=13\,$TeV~\cite{ATLAS-CONF-2017-027}. 
The production cross-section was calculated at NLO in the 5-flavour scheme using {\tt MadGraph-2.5.4}~\cite{1405.0301}. 
The current limits only provide meaningful constraints for $Z_{BL3}$ masses $\lesssim2\,$TeV. 
Hence, unlike many previous models, one can comfortably account for the LFU anomalies with a $U(1)_{(B-L)_3}$ gauge coupling that remains perturbative up to the Planck scale. 

\subsection{Heavy Fermions}
The vector-like fermions that generate the quark and lepton mass matrices may also have observable low-energy consequences. 
In particular, they will modify the couplings of the SM fermions to the usual $Z$ boson. 
Integrating out the heavy fermion leads to the effective operator involving the first two families of leptons
\begin{align}
\bar l_L H\frac{Y_E Y_E^\dagger}{M_E^2} i\gamma^\mu (D_\mu H^\dagger) l_L\,.
\end{align}
LEP measurements strongly constrain such operators, which induce lepton universality violation in the couplings of the $Z$ boson and lead to bounds~\cite{Agashe:2014kda}:
\begin{align}\label{LEPbd}
\frac{v^2|Y_E|^2}{2M_E^2}c_{\theta_l}^2&<  1.0\times 10^{-3}\,, \\
\frac{v^2|Y_E|^2}{2M_E^2}s_{\theta_l}^2&<  6.1\times 10^{-4}\,.
\end{align}
There are similar bounds on $|Y_D|^2/M_D^2$ from modifications of the $Z\bar{b}b$ coupling~\cite{1508.07010}.
Let us stress here that the effects induced by fermions are not in general correlated with either $Z_{BL3}$ couplings or SM masses and mixings, the latter depending on the combination of parameters $Y_EY_E^\prime\phi_l/M_E$.
In particular a small $Y_E$ ($Y_D$) such that the bound in Eq.~(\ref{LEPbd}) is satisfied while $E$ ($D$) being relatively light is a possibility.

In this sense these heavy fermions could  potentially be within LHC reach. 
Current searches are sensitive to vector-like quarks with masses of order 1 TeV~\cite{ATLAS-CONF-2017-015, ATLAS-CONF-2016-102, 1509.04261}.

\begin{figure}[htbp]
  \centering
  \includegraphics[width=0.35\textwidth]{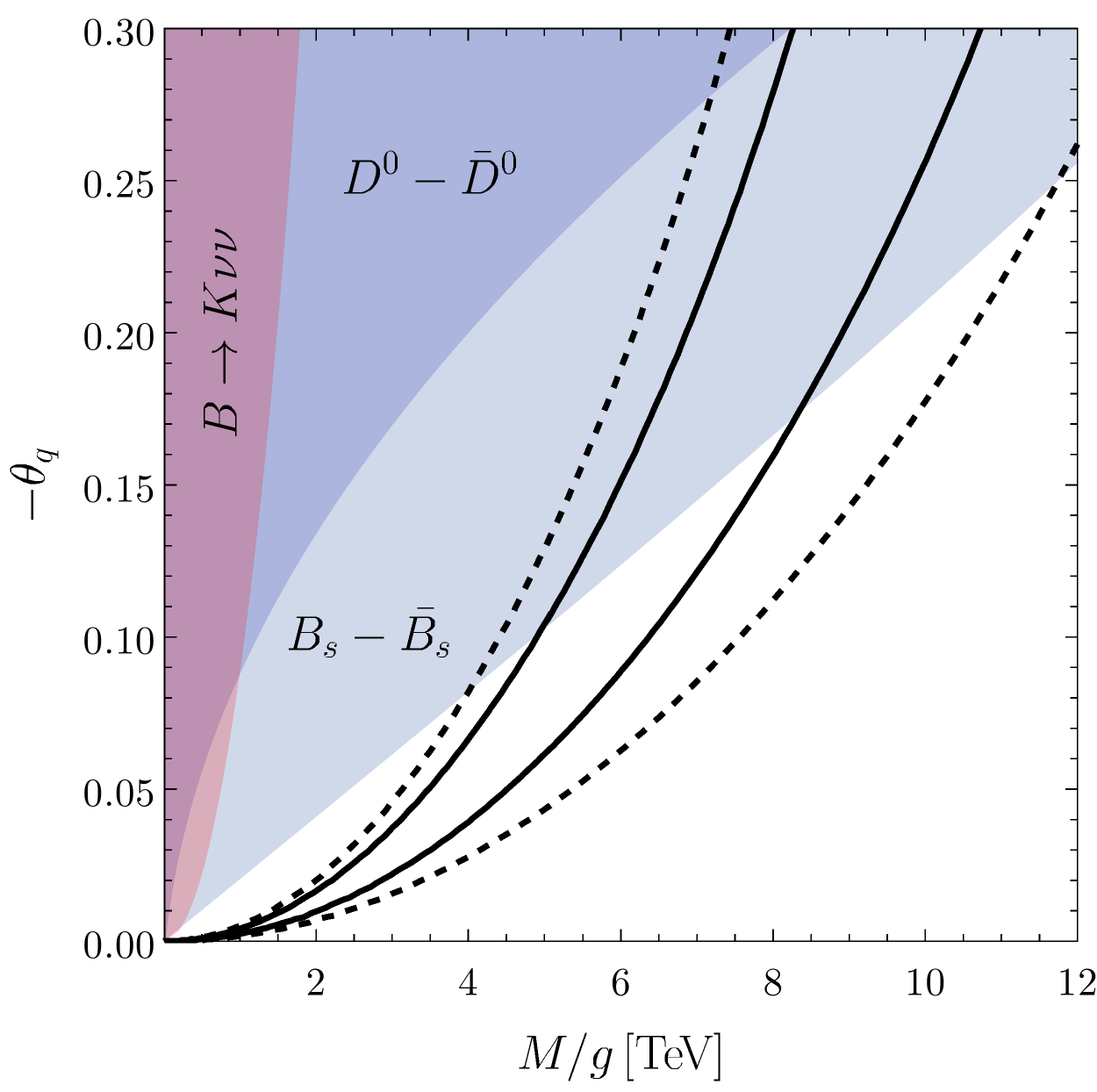} 
  \caption{Best-fit region to the LFU anomalies at $1\sigma$ (solid lines) and $2\sigma$ (dashed lines). The shaded regions are excluded by existing measurements at 95\% CL. We have fixed $\theta_l=\pi/2$.}
  \label{fig:M-theta_q}
\end{figure}

\begin{figure}[htbp]
  \centering
  \includegraphics[width=0.35\textwidth]{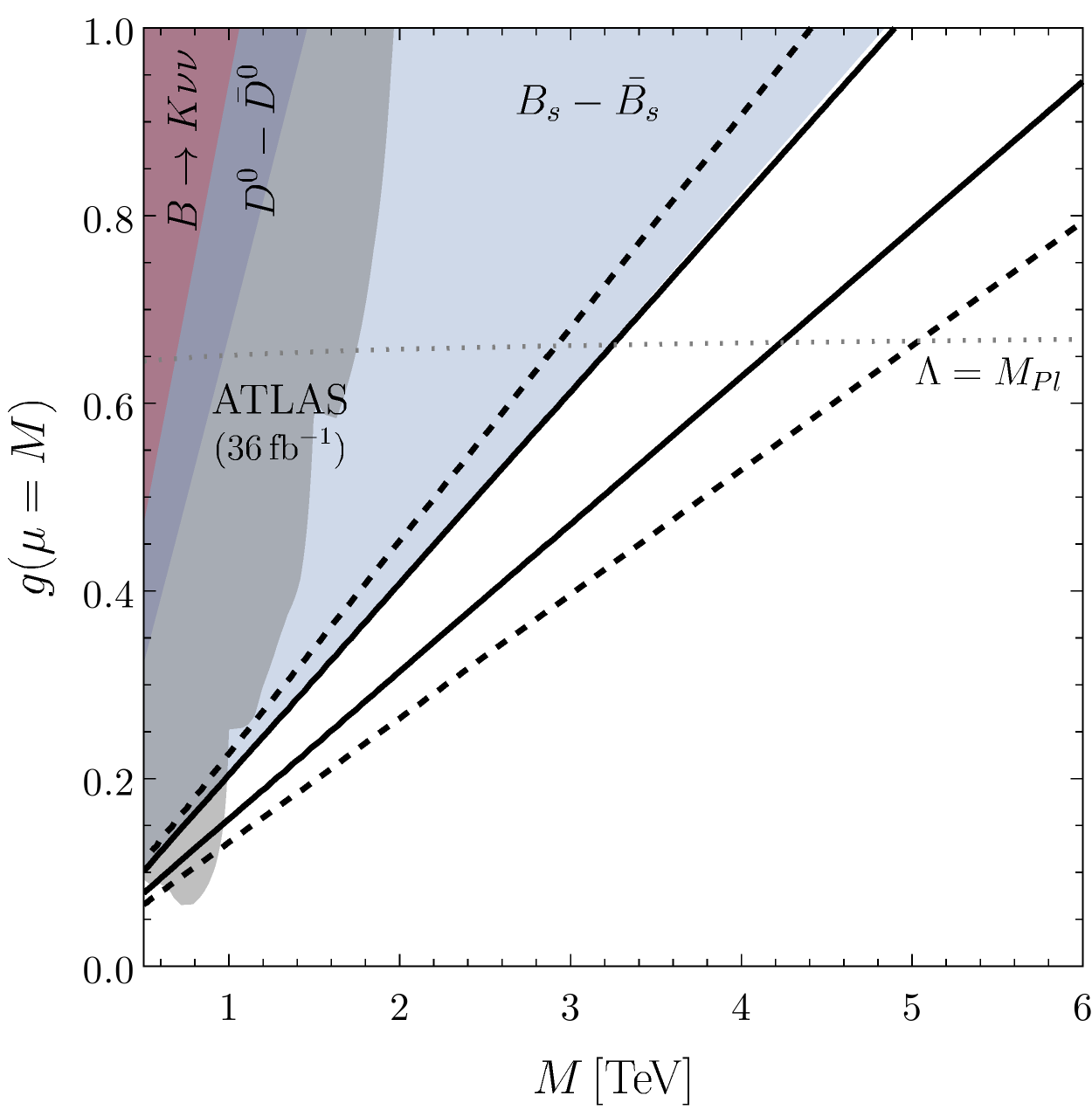} 
  \caption{Same as Fig.~\ref{fig:M-theta_q}, but shown in the $M-g$ plane. The dotted line shows the value of the coupling that runs to a Landau pole at the Planck scale. We have fixed $\theta_q=-0.1$.}
  \label{fig:M-g}
\end{figure}

\section{Outlook}
As we pointed out in our previous paper~\cite{1704.08158}, $SU(3)_Q\times SU(3)_L \times U(1)_{B-L}$ is the largest anomaly-free local symmetry that can be added within the SM+3$\nu_R$. 
Furthermore any anomaly-free, vector-like local U(1) symmetry is one of the subgroups of $SU(3)_Q \times SU(3)_L\times U(1)_{B-L}$. 
If we adopt two requirements: (i) at least two right-handed neutrinos should have super-heavy Majorana masses for leptogenesis; and (ii) we have sufficient suppression of $K^0-{\bar K^0}$ and $D^0-{\bar D^0}$ oscillations, we end up with only two classes of vectorial $U(1)$'s at the TeV scale:
\begin{equation}
  T_Q=\rm{diag}\left(\frac{1}{9}+a, \frac{1}{9}+a, \frac{1}{9}-2a\right)\!, \,\,\, T_L= \rm{diag}(0,0,-1) ,
\end{equation}
and
\begin{equation}
  T_Q=\rm{diag}(a, a, -2a), \quad T_L= \rm{diag}(0,1,-1) \,.
\end{equation}

The flavoured $B-L$ we have considered in this letter is a special case of the first class ($a=-1/9$).\footnote{The second class of models was considered as an explanation for the LFU anomalies in~\cite{1503.03477}. For a discussion of other models motivated by anomaly cancellation see Ref.~\cite{1705.03447}.} 
It is also the unique choice which minimises the LHC constraints, thus allowing the $U(1)_{(B-L)_3}$ gauge coupling to remain perturbative up to the Planck scale, while simultaneously explaining the LFU anomalies. 
This fact also encourages us to consider another fascinating unification framework, that is, $SO(10)_1 \times SO(10)_2 \times SO(10)_3$. 
Here, quarks and leptons in the $i$th generation belong to the individual $SO(10)_i$ GUT. 
We then assume a breaking near the Planck scale $(SO(10))^3 \rightarrow\mathcal{G}_{SM}\times U(1)_{(B-L)_3}$, with $\mathcal{G}_{SM}$ the gauge group in the SM.

Finally, if one considers chiral $U(1)$ symmetries there are many more possibilities consistent with anomaly cancellation. 
Let us briefly comment on one particularly interesting case: flavoured $U(1)_{5\rm{ness}}$ in GUTs. 
This is the unique chiral $U(1)$ local symmetry that satisfies the two conditions above and is consistent with $SU(5)$. 
Here, the $U(1)_{5\rm{ness}}$ charges of the quarks and leptons in the third generation are $\ell_L(-3)$, $d_R^c(-3)$, $q_L(+1)$, $u_R^c(+1)$, $e_R^c(+1)$ and $\nu_R^c(+5)$. 
Since the first and second generations have vanishing $U(1)_{5\rm{ness}}$ charges, this symmetry is not equivalent to $U(1)_{(B-L)_3}$. 
The low-energy phenomenology will be similar to that discussed in Section~\ref{sec:pheno}.

\appendix

\section{Formulae for the rotation to the mass basis} \label{appendix}
Here we detail the connection between the model parameters and the SM fermion masses and mixings, together with the explicit conditions to reproduce the ansatz in Eq.~(\ref{Rots}).

One can take $\hat Y_d=\mathbf V_L^d \mbox{diag}\left(y_d\,,y_s\right)  (\mathbf V_R^d)^\dagger$ in full generality and, after expanding in $\epsilon \sim \phi_q/(MY_b), \hat Y_d/Y_b$, one has $y_b=Y_b+\mathcal O(\epsilon)$ and mixing matrices:
\begin{align}
  U_{d_L}=&\left(\begin{array}{cc}
  \mathbf V_L^d& -\frac{Y_D'\phi^*_q}{M_DY_b}Y_D \nonumber\\
  \frac{Y_D^{\prime*}\phi_q}{M_D^*Y_b^*}Y_D^\dagger \mathbf V_L^d  &1\\
  \end{array}\right) +\mathcal{O}(\epsilon^2) \,, \\
  U_{d_R}=&\left(\begin{array}{cc}
  \mathbf V_R^d &-\frac{Y_Q^{\prime*}\phi_q^*}{M_Q
  ^*Y_b^*}Y_Q^*\\
  \frac{Y_Q^\prime \phi_q}{M_QY_b} Y_Q^T \mathbf V_R^d  &1\\
  \end{array}\right) +\mathcal{O}(\epsilon^2) \,.
\end{align}
In particular, note that for $M_Q\gg M_{D,U}$ RH  third generation quark mixing is even further suppressed. 
The above relations translate to up-type quarks with the obvious substitutions. 
The Cabibbo-Kobayashi-Maskawa matrix reads:
\begin{widetext}
\begin{align} \label{CKM}
  V_{CKM}=(U_L^u)^\dagger U_{L}^d=
  \left(\begin{array}{cc}\left(\mathbf V_L^u\right)^\dagger \mathbf V_L^d &
  \left(\mathbf V_L^u\right)^\dagger\left(\frac{Y_U'\phi^*_q}{M_UY_t}Y_U-\frac{Y_D'\phi^*_q}{M_DY_b}Y_D\right) \\ 
  \left(\frac{Y_D^{\prime *}\phi_q}{M_D^*Y_b^*}Y_D^\dagger-\frac{Y_U^{\prime *}\phi_q}{M_U^*Y_t^*}Y_U^\dagger\right)\mathbf V_L^d&1
  \end{array} \right)+\mathcal O(\epsilon^2) \,.
\end{align}
\end{widetext}
This implies a $(B-L)_3$ current for down-type LH quarks:
\begin{align}
U_{d_L}^\dagger  T^q U_{d_L}=\frac13\left(\begin{array}{cc}
  0
  &
  \frac{Y_D'\phi_q^*}{M_DY_b}\mathbf V_L^{d\dagger} Y_D \\
  \frac{Y_D^{\prime*}\phi_q}{M_D^*Y_b^*}Y_D^\dagger \mathbf V_L^d  &1\\
  \end{array}\right)+\mathcal O\left(\epsilon^2
  \right) ,
\end{align}
with $T^q$ as in Eq.~(\ref{Torig}). Analogous relations hold for for LH up-type  and up and down-type RH quarks. 
The ansatz in Eq.~(\ref{Rots}) can then be obtained choosing $(\mathbf V_{L}^d)^\dagger\lambda_D=(0,\lambda_D)^T$
so that $\theta_q$ is confined to the $2-3$ sector, 
the mixing matrix $U_{u_L}$ can be found by inverting Eq.~(\ref{CKM}) and finally
for the RH quarks the limit $M_Q\gg M_{D,U}$ yields $U_{(d,u)_{R} }^\dagger T^q U_{(d,u)_R}\simeq\, T^q$.
 
 In the charged lepton sector we have
 \begin{align}
  &\left(\begin{array}{cc}\bar l_L& \bar {\mathrm l}_L^3 \end{array}\right)
  H \left(\begin{array}{cc}
  \hat Y_e &   -\frac{Y_E'\phi_l}{M_E} Y_E   \\
  - \frac{Y_L^\prime\phi_l^*}{M_L}Y_L^T
  & Y_\tau \\
  \end{array} 
  \right)
  \left(\begin{array}{c} e_{R}\\  \tau_R\\\end{array}\right) .
\end{align}
If one assumes $\hat Y_e=\,$diag$(y_e\,,Y_\mu)$ and $\lambda_E=(0,\lambda_E)^T$, and expands in $\epsilon \ll 1$ with $\epsilon\sim  \hat Y_e/Y_\tau \,,\,Y_L^\prime Y_L\phi_l^*/M_L Y_\tau$, (but note that we do not expand in $ Y_E^\prime Y_E\phi_l/M_E Y_\tau$) the diagonalisation yields:
\begin{align}
  y_\tau^2&=Y_\tau^2+ \left|\frac{Y_E'\phi_l Y_E }{M_E} \right|^2+\mathcal O(\epsilon^2) 
   \end{align}\begin{align}
  U_L^e&=\left( \begin{array}{ccc}
  1&0&0\\
  0&c_{\theta_l} & s_{\theta_l}\\
  0&-s_{\theta_l^\dagger}&c_{\theta_l}\\
  \end{array}
  \right) +\mathcal O(\epsilon),
\end{align}
with $\tan(\theta_l)=-Y_E' Y_E \phi_l / (M_EY_\tau)$, whereas $U_{e_R}=1+\mathcal O(\epsilon)$ and $y_\mu, y_e$ are order $\epsilon$. 

Finally the neutrino mass matrix has all entries of the same order and it is not simple to give the unitary matrix $U_{\nu_L}$ for arbitrary $3\times 3$ matrices
$Y_\nu$\,,$M_{\nu_R}$. 
Nevertheless the number of parameters is large enough that we can assume an $U_{\nu_L} $ as in Eq.~(\ref{Rots}) which already incorporates $U_{PMNS}$.

\noindent {\bf Note added:} During the completion of this work Ref.~\cite{1705.00915} appeared on the arXiv and considers a similar explanation for the anomalies.

\noindent {\bf{Acknowledgements}}
This work is supported by Grants-in-Aid for Scientific Research from the Ministry of Education, Culture, Sports, Science, and Technology (MEXT), Japan, No. 26104009 (T.T.Y.), No. 16H02176 (T.T.Y.) and No. 17H02878 (T.T.Y.), and by the World Premier International Research Center Initiative (WPI), MEXT, Japan (P.C., C.H. and T.T.Y.). 
This project has received funding from the European Union's Horizon 2020 research and innovation programme under the Marie Sk\l{}odowska-Curie grant agreement No 690575. 

\bibliography{paper}

\end{document}